\documentclass[12pt,a4paper]{article}
\usepackage{amsfonts}
\usepackage{amssymb}
\usepackage{latexsym}
\usepackage{graphicx}

\begin{document}
\begin{center}
{\Large \bf On a stabilized warped brane world\\ without Planck brane} \\

\vspace{4mm}

Mikhail N. Smolyakov, Igor P. Volobuev\\ \vspace{0.5cm} Skobeltsyn
Institute of Nuclear Physics, Moscow State University \\ 119991
Moscow, Russia\\
\end{center}

\begin{abstract}
We discuss a stabilized brane world model with two branes,
admitting the solution to the hierarchy problem due to the warped
extra dimension and possessing a remarkable feature: the strength
of gravitational interaction is of the same order on both branes,
contrary to the case of the Randall-Sundrum model with a
hierarchical difference of gravitational strength on the branes.
The solution also admits the existence of two branes with an equal
strength of gravitational interaction, which is of interest for
treating  the matter on the "mirror" brane as dark matter.
\end{abstract}

\section{Introduction}
Since in papers \cite{Gogberashvili:1998vx,Randall:1999ee} it was
shown that theories with warped extra dimension could solve the
hierarchy problem, such theories are widely discussed in the
literature (see, for example, reviews
\cite{Rubakov:2001kp,Kubyshin:2001mc}). It turned out that the RS1
model \cite{Randall:1999ee} includes the massless radion, which
contradicts the experimental data even at the classical level, and
thus this model demands a stabilization. The first method was
proposed in paper \cite{Goldberger:1999un}, were the size of the
extra dimension is defined by the minimum of the effective
potential of the five-dimensional scalar field. At the same time
the back reaction of the scalar field on the background metric is
not taken into account by this mechanism. This problem was solved
in the model proposed in \cite{DeWolfe}. It is necessary to note
that in the method proposed in \cite{DeWolfe} the size of the
extra dimension is defined not by the minimum of the effective
four-dimensional scalar field potential, but by the boundary
conditions on the branes. With the proper choice of the model
parameters it is possible to obtain the background solution for
the metric which is close to the original Randall-Sundrum solution
\cite{our}. It can be considered as the stabilized Randall-Sundrum
model.

The main feature of the Randall-Sundrum model is the existence of
two branes, which differ significantly in the strength of the
four-dimensional gravity. Indeed, the four-dimensional Planck
masses, defined by the coupling constant of massless
four-dimensional graviton, are \cite{Rubakov:2001kp,Boos:2002ik}
$$M_{Pl}^{2}=\frac{M^{3}}{k}\left(e^{2kL}-1\right)$$
for the $TeV$ brane and
$$M_{Pl}^{*2}=\frac{M^{3}}{k}\left(1-e^{-2kL}\right)$$
for the Planck brane, where $M$ is the five-dimensional Planck
mass, $k$ is the inverse anti-de Sitter radius, $L$ is the size of
the extra dimension, $M\simeq k\sim 1 TeV$ and $kL\approx 36$.
Obviously, for this choice of the parameters $M_{Pl}\sim
10^{16}\,TeV$, whereas $M_{Pl}^{*}\sim 1\,TeV$. Thus, the branes
are very different from the gravitational point of view, which
leads to very different physics on both branes. There arises a
question, whether it is possible to construct a stabilized brane
world model with branes having comparable (or even equal) strength
of effective four-dimensional gravity.

In this short paper we discuss a stabilized brane world model,
based on the background solution presented in
\cite{Kakushadze:2000zp,Flanagan:2001dy}. Although the solution is
quite known, we found that it possesses an interesting property,
when applied to compact extra dimension. Namely, it allows one to
obtain any values of the four-dimensional Planck masses with
respect to each other, retaining the main advantages of warped
brane worlds -- strong five-dimensional gravity and the solution
to the hierarchy problem. In other words, there can be two $TeV$
(SM) branes or even the case with one $TeV$ brane and one brane
with the gravity much weaker than that on the $TeV$ brane (the
notations "Planck brane" and "$TeV$ brane" are used in the same
sense as in the Randall-Sundrum model -- the Planck brane is the
one with the stronger gravity, whereas the $TeV$ brane is the
brane with the weaker gravity).

\section{The model}
To start with, let us denote the coordinates  in five-dimensional
space-time $E=M_4\times S^{1}$  by $\{ x^N\} \equiv
\{x^{\mu},y\}$, $N= 0,1,2,3,4, \, \mu=0,1,2,3 $, the coordinate
$x^4 \equiv y, -L\leq y \leq L$ parameterizing the fifth dimension
with identified points $-y$ and $y$. The branes are located at the
points $y=0$ and $y=L$.

The action of the stabilized brane world model can be written as
\begin{eqnarray}\label{actionDW}
S= \int d^{4}x \int_{-L}^L dy \sqrt{-g} \left[  2 M^3R
-\frac{1}{2} g^{MN}\partial_M\phi\partial_N\phi-V(\phi)\right] -\\
\nonumber -\int_{y=0}\sqrt{-\tilde
g}\lambda_{1}(\phi)d^{4}x-\int_{y=L}\sqrt{-\tilde
g}\lambda_{2}(\phi)d^{4}x,
\end{eqnarray}
Here $V(\phi)$ is a bulk scalar field potential and
$\lambda_{i}(\phi)$, $i=1,2$, are the brane scalar field
potentials, $\tilde{g}=det\tilde g_{\mu\nu}$, and $\tilde
g_{\mu\nu}$  denotes the metric induced on the brane. The
signature of the metric $g_{MN}$ is chosen to be $(-,+,+,+,+)$.

The standard ansatz  for the  metric and the scalar field, which
preserves the Poincar\'e invariance in any four-dimensional
subspace $y=const$, looks like
\begin{equation}\label{metricDW}
ds^2=  e^{-2A(y)}\eta_{\mu\nu}  {dx^\mu  dx^\nu} +  dy^2 \equiv
\gamma_{MN}(y)dx^M dx^N, \quad \phi(x,  y) = \phi(y),
\end{equation}
 $\eta_{\mu\nu}$ denoting the flat Minkowski metric.

For this ansatz the Einstein and the scalar field equations
derived from action (\ref{actionDW}) reduce to the following
system:
\begin{eqnarray}\label{yd}
&\frac{dV}{d\phi}+\frac{d\lambda_1 }{d\phi}\delta(y)
+\frac{d\lambda_2 }{d\phi}\delta(y-L)= -4A'\phi'+\phi''&\\
\nonumber &12M^3(A')^2+\frac{1}{2}(V-\frac{1}{2} (\phi')^2)=0& \\
\nonumber
&\frac{1}{2}\left(\frac{1}{2}(\phi')^2+V+\lambda_1\delta(y)+\lambda_2\delta(y-L)
\right)=-2M^3\left(-3A''+6(A')^2\right).&
 \end{eqnarray}

Let us consider a special class of potentials, which can be
represented as
$$
V(\phi)=\frac{1}{8} \left(\frac{d
W}{d\phi}\right)^2- \frac{1}{24M^3}W^2(\phi).
$$
Then the solutions of the first order differential equations
\begin{equation}\label{sol}
\phi'(y) =sign(y)\frac{1}{2} \frac{dW}{d\phi}, \quad A'(y) =
sign(y)\frac{1}{24M^3}W(\phi)
\end{equation}
solve equations (\ref{yd}) in the bulk  \cite{DeWolfe,Brandhuber}.

 Let us consider a linear  function $W(\phi)$ as suggested in papers
 \cite{Kakushadze:2000zp,Flanagan:2001dy}:
\begin{equation}\label{W}
W(\phi)=\alpha\phi, \quad
V=\frac{\alpha^{2}}{8}-\frac{\alpha^{2}}{24M^{3}}\phi^{2}.
\end{equation}
Using  equations (\ref{sol}) one gets the corresponding background
solution, which looks like
\begin{equation}\label{backgr}
\phi=\frac{\alpha}{2}|y|-\frac{\alpha L_{1}}{4},\quad
A=\frac{\alpha^{2}}{96M^{3}}\left[\left(|y|-\frac{L_{1}}{2}\right)^{2}+C\right],
\end{equation}
where $L_{1}$ and $C$ are integration constants treated as
parameters.

In order the equations of motion be valid on the branes too, one
can take the brane potentials $\lambda_{i}(\phi)$, $i=1,2$, in the
form
\begin{eqnarray}\label{bound1}
\lambda_{1}(\phi)=W(\phi)+\beta_{1}^{2}\left(\phi-\phi_{1}\right)^{2},\\
\label{bound2}
\lambda_{2}(\phi)=-W(\phi)+\beta_{2}^{2}\left(\phi-\phi_{2}\right)^{2}.
\end{eqnarray}
It is easy to check that the equations of motion are satisfied
provided
\begin{eqnarray}
\phi|_{y=0}=\phi_{1},\\
\phi|_{y=L}=\phi_{2},
\end{eqnarray}
which means that
\begin{eqnarray}
L_{1}=-\frac{4\phi_{1}}{\alpha},\\
L=\frac{2(\phi_{2}-\phi_{1})}{\alpha}
\end{eqnarray}
(we suppose that $\phi_{1}<0$, i.e. $L_{1}>0$). Thus, we see that
the size of the extra dimension is fixed. The parameters of the
potentials $\alpha, \phi_{1,2}, \beta_{1,2}$, when made
dimensionless by the fundamental five-dimensional energy scale of
the theory $M$, do not contain a hierarchical difference. We note
again that fixation of the size of the extra dimension is caused
by the boundary conditions on the branes unlike the case discussed
in \cite{Goldberger:1999un}, where the size of the extra dimension
is defined by the minimum of an effective four-dimensional  scalar
field potential.

Let us suppose that we live on the brane at $y=L$ and $L\le
L_{1}$. In order to have Galilean four-dimensional coordinates on
this brane \cite{Rubakov:2001kp,Boos:2002ik}, we choose the warp
factor such that $e^{-2A}|_{y=L}=1$, i.e.
$C=-\left(L-\frac{L_{1}}{2}\right)^{2}$. Since the wave function
of the massless tensor graviton in the extra dimension $\sim
e^{-2A}$ \cite{our}, a standard technique (see, for example,
\cite{Kubyshin:2001mc} for details) gives us an expression for the
four-dimensional Planck mass on our brane
\begin{eqnarray}\label{Plmass0}
&M_{Pl}^{2}=M^{3}\int_{-L}^{L}e^{-2A}dy=
M^{3}2e^{\frac{\alpha^{2}(2L-L_{1})^{2}}{192M^{3}}}\int_{0}^{L}e^{-\frac{\alpha^{2}}{48M^{3}}\left(y-\frac{L_{1}}{2}\right)^{2}}dy= &\\
\nonumber &=
M^{3}2e^{\frac{\alpha^{2}(2L-L_{1})^{2}}{192M^{3}}}\int_{-\frac{L_{1}}{2}}^{\frac{2L-L_{1}}{2}}e^{-\frac{\alpha^{2}}{48M^{3}}y^{2}}dy\simeq
M^{3}2e^{\frac{\alpha^{2}(2L-L_{1})^{2}}{192M^{3}}}\int_{-\infty}^{\infty}e^{-\frac{\alpha^{2}}{48M^{3}}y^{2}}dy=&\\
\nonumber &=M^{3}\frac{2\sqrt{48\pi
M^{3}}}{\alpha}e^{\frac{\alpha^{2}(2L-L_{1})^{2}}{192M^{3}}}&
\end{eqnarray}
and
\begin{eqnarray}\label{Plmass}
M_{Pl}\approx
M\frac{5M^{\frac{5}{4}}}{\sqrt{\alpha}}\,e^{\frac{\alpha^{2}(2L-L_{1})^{2}}{384M^{3}}}.
\end{eqnarray}
Let us suppose that all fundamental parameters of the theory lie
in the $TeV$ range. To have the hierarchy problem solved, i.e. to
have $M_{Pl}\sim 10^{16} TeV$, one should take
\begin{equation}
\frac{\alpha
(2L-L_{1})}{M^{\frac{3}{2}}}=\frac{4\phi_{2}}{M^{\frac{3}{2}}}\approx
120.
\end{equation}
With these values of the parameters (and if $L$ and $L_{1}$ are of
the same order), the approximation used in (\ref{Plmass0}) is very
good. Also note that the four-dimensional Planck mass
(\ref{Plmass}) depends mainly on $\phi_{2}$, not $\phi_{1}$.

\begin{figure}
\centering
\includegraphics[width=16cm,height=9cm]{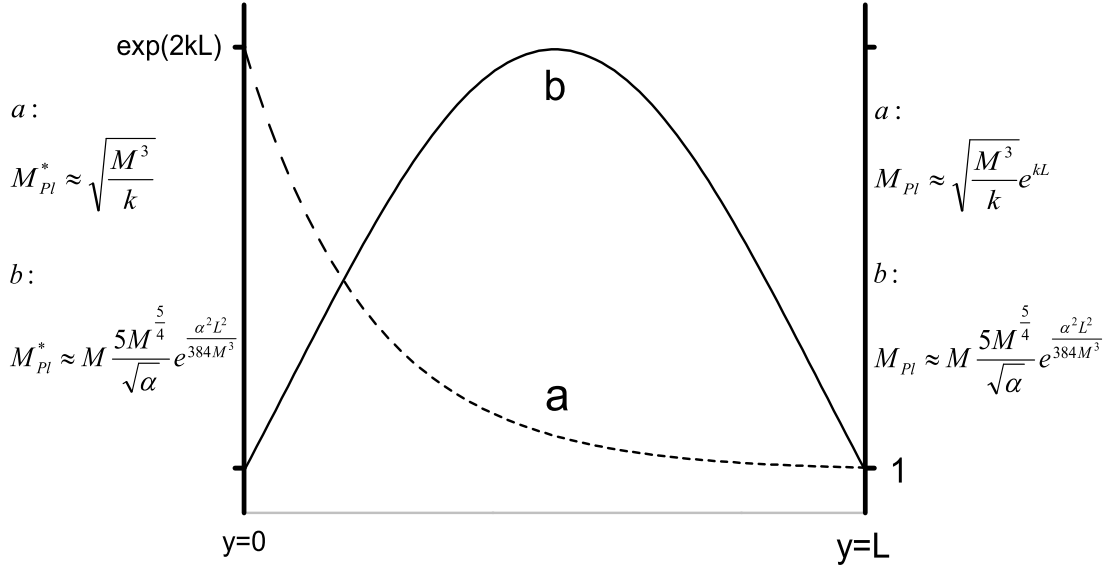}
\caption{Warp factors in the Randall-Sundrum model (dashed line,
$a$) and in the model under discussion (solid line, $b$). The
branes are located at the points $y=0$, $y=L$ of the extra
dimension. The formulas for four-dimensional Planck masses in
Galilean coordinates on each brane for both cases are presented
(for case $b$ we consider $L_{1}=L$). All the parameters are
supposed to lie in the $TeV$ range, the hierarchy problem is
solved due to the exponential factors $e^{kL}\simeq
e^{\frac{\alpha^{2}L^{2}}{384M^{3}}}\sim 10^{16}$.}
\label{warpfactor}
\end{figure}

Now let us discuss some properties of this solution. On the second
brane
$$e^{-2A}|_{y=0}=e^{-\frac{\alpha^{2}}{48M^{3}}\left[L\left(L_{1}-L\right)\right]}<1$$
for $L<L_{1}$. Let us compare this behavior of the warp factor
with that in the RS1 model. Indeed, in the RS1 model (as well as
in the stabilized case \cite{DeWolfe}), if we live on the $TeV$
brane, there exists the Planck brane, where gravity is much
stronger. In the case of background solution (\ref{backgr}) the
gravity on our brane is weak in comparison with the bulk gravity
strength, but the gravity on the brane at $y=0$ is even weaker.
Thus, we have no Planck brane as a physical object in this model.
In the RS1 model the exponent $A(y)$ takes its smallest value at
the point, where the Planck brane is located, whereas in the case
of solution (\ref{backgr}) the analogous point $y=L_{1}/2$ lies in
the bulk, i.e. the massless tensor graviton, whose wave function
in the extra dimension is proportional to $e^{-2A}$, is localized
in the bulk (for the case $L_{1}=L$ see Figure~\ref{warpfactor}).
Such situation can lead to  interesting consequences for the
effect of "mirror" matter (located on the "mirror" brane) on our
brane and for the case of universal extra dimensions.

It is not difficult to calculate the four-dimensional Planck mass
on the brane at $y=0$. To this end we pass to Galilean
four-dimensional coordinates on that brane, which means that the
integration constant in solution (\ref{backgr}) for $A(y)$ should
be taken such that $A(y)|_{y=0}=0$, i.e.
\begin{eqnarray}\label{backgrL}
A=\frac{\alpha^{2}}{96M^{3}}\left[\left(y-\frac{L_{1}}{2}\right)^{2}-\frac{L_{1}^{2}}{4}\right].
\end{eqnarray}
Carrying out the calculations analogous to those presented above,
we obtain
\begin{eqnarray}\label{PlmassL}
M^{*}_{Pl}\approx
M\frac{5M^{\frac{5}{4}}}{\sqrt{\alpha}}\,e^{\frac{\alpha^{2}L_{1}^{2}}{384M^{3}}}.
\end{eqnarray}
Thus
\begin{eqnarray}\label{PlmassRel}
M^{*}_{Pl}=M_{Pl}\,e^{\frac{\alpha^{2}L\left(L_{1}-L\right)}{96M^{3}}}.
\end{eqnarray}
Note that $M_{Pl}$ and $M^{*}_{Pl}$ were calculated in
four-dimensional coordinates, which are Galilean on the branes at
$y=L$ and $y=0$ respectively, not in a four-dimensional coordinate
system, which is common for both branes.

A quite peculiar case is $L=L_{1}$. For this choice of the
parameters, the warp factor has equal values on both branes, and
we have two equal branes from the gravitational point of view! At
the same time the hierarchy problem is solved in this case also
because of the quadratic behavior of the function $A(y)$ and the
corresponding behavior of the warp factor, see
Figure~\ref{warpfactor}. The peculiar feature of the model is that
the hierarchy problem appears to be solved for both branes,
contrary to the case of the RS1 model, in which the hierarchy
problem is solved only for brane at $y=L$.

The branes in this case are  not only gravitationally equivalent,
but also have the same negative  tension (energy density), which
is  characteristic of the TeV brane in the RS1 model. If not only
the gravitational constants on both branes are of the same order,
but  the density of the "mirror" matter is also of the same order
as that of the ordinary matter on our brane, the "mirror" matter
of WIMP type may play, in principle, the role of dark matter. This
possibility deserves a more detailed investigation.

Now let us discuss stability of this background solution under
small fluctuations of the fields. To this end we should consider
the linearized theory. The physical degrees of freedom in
five-dimensional brane world models stabilized by the scalar field
were described in \cite{our} for the general case of stabilizing
scalar field potential. It has been shown that if there exists a
background solution of form (\ref{metricDW}) to field equations
(\ref{yd})  and if the size of the extra dimension is fixed by
boundary conditions on the branes, the tensor sector of
Kaluza-Klein excitations does not contain tachyons or fields with
the wrong sign of the kinetic term (ghosts). As for the scalar
sector, it does not contain tachyons, ghosts and massless (from
the four-dimensional point of view) modes, if \cite{our}
\begin{equation}\label{parconstr}
\left(\frac{1}{2}\frac{d^2 \lambda_1}{d \phi^2} -
\frac{\phi''}{\phi'}\right)|_{y=0+\epsilon}>0, \quad
\left(\frac{1}{2}\frac{d^2\lambda_2}{d \phi^2} +
\frac{\phi''}{\phi'}\right)|_{y=L-\epsilon}>0.
\end{equation}
Note that conditions (\ref{parconstr}) do not involve the bulk
potential $V(\phi)$, and it may be unbounded from below as in
equation (\ref{W}).

One can easily find that for the scalar field configuration
satisfying equation (\ref{sol}) and potentials given by
(\ref{bound1}), (\ref{bound2}) conditions (\ref{parconstr}) reduce
to $\beta_{1,2}^{2}>0$ and are satisfied. Thus, the model under
consideration is indeed stable, at least perturbatively.

It is also worth mentioning that in the case of gravitationally
equivalent branes there exists another interesting possibility to
stabilize the size of the extra dimension. There is a good reason
to guess that the branes are equivalent initially, i.e. the scalar
field potentials on the branes are of the same form. Namely, the
(fine-tuned) brane potentials can be chosen to be non-polynomial
\begin{equation}\label{eqlambdas}
\lambda_{1,2}(\phi)=-3\left(\frac{\rho\alpha^{2}}{4}\right)^{\frac{1}{3}}+\frac{\rho}{\phi^{2}}
\end{equation}
with $\rho>0$. From the boundary conditions on the branes, which
follow from equations (\ref{yd}), one can easily obtain
\begin{equation}
L=L_{1}=4\left(\frac{2\rho}{\alpha^{4}}\right)^{\frac{1}{3}}.
\end{equation}
The stability conditions (\ref{parconstr}) are  fulfilled for the
choice of the potentials in (\ref{eqlambdas}).

As for the wave functions, coupling constants and masses of the
tensor and scalar Kaluza-Klein modes, it seems that it is
impossible to solve the corresponding equations of motion
analytically for the case of background solution (\ref{backgr})
 (except for the tensor zero mode, which is proportional
to $e^{-2A}$). But it is quite obvious that masses of the lowest
excitations are of the order of  $ 1/L$ (since the model is
stabilized, there is no massless radion), and the corresponding
coupling constants should  also be expressed through the
fundamental parameters of the theory. Thus, all these parameters
should lie in the $TeV$ energy range, as it usually happens in the
brane world models.

\section{Conclusion}
In this paper we have discussed the stabilized brane world model,
admitting the solution to the hierarchy problem on both branes,
contrary to the case of the Randall-Sundrum-type models, in which
solution to the hierarchy problem can be obtained only for one
brane. The stability of the model under fluctuations of metric and
scalar field is provided by the fulfilment of conditions
(\ref{parconstr}), which are valid for any brane world model with
the action of the form (\ref{actionDW}), stabilized by a scalar
field \cite{our}. We also show that if one assumes the branes to
have an equal structure, namely, the brane potential to be the
same on both branes, there also exists a solution with fixed size
of the extra dimension. In this case the effective
four-dimensional Planck masses on both branes appear to be equal.

\section*{Acknowledgments}

The work was supported by grant of Russian Ministry of Education
and Science NS-1456.2008.2. M.S. acknowledges support of grant for
young scientists MK-5602.2008.2 of the President of Russian
Federation, grant of the "Dynasty" Foundation and scholarship for
young scientists of Skobeltsyn Institute of Nuclear Physics of
M.V. Lomonosov Moscow State University.

\end{document}